# Structural evolution of epitaxial SrCoO$_x$ films near topotactic phase transition

**Authors:** Hyoungjeen Jeen[1,2] and Ho Nyung Lee[1]*

**Affiliations:**

[1]Materials Science and Technology Division, Oak Ridge National Laboratory, Oak Ridge, Tennessee 37831, USA.

[2]Department of Physics, Pusan National University, Busan, 609735, South Korea.

*To whom correspondence should be addressed. E-mail: hnlee@ornl.gov

**Abstract:** Control of oxygen stoichiometry in complex oxides via topotactic phase transition is an interesting avenue to not only modifying the physical properties, but utilizing in many energy technologies, such as energy storage and catalysts. However, detailed structural evolution in the close proximity of the topotactic phase transition in multivalent oxides has not been much studied. In this work, we used strontium cobaltites (SrCoO$_x$) epitaxially grown by pulsed laser epitaxy (PLE) as a model system to study the oxidation-driven evolution of the structure, electronic, and magnetic properties. We grew coherently strained SrCoO$_{2.5}$ thin films and performed post-annealing at various temperatures for topotactic conversion into the perovskite phase (SrCoO$_{3-\delta}$). We clearly observed significant changes in electronic transport, magnetism, and microstructure near the critical temperature for the topotactic transformation from the brownmillerite to the perovskite phase. Nevertheless, the overall crystallinity was well maintained without much structural degradation, indicating that topotactic phase control can be a useful tool to control the physical properties repeatedly via redox reactions.

Epitaxial oxygen sponges are a revitalized class of materials since they can instantaneously transform the crystal structure at temperatures as low as 200 °C.[1] This structural transformation accompanies changes in the crystal symmetry and chemical composition by losing or gaining a constituent element.[2-7] Due to its reversibility of the crystal structure and stoichiometry, complex oxides, which undergo a topotactic phase transition, are proposed as promising energy materials for use in batteries and solid oxide fuel cells. In addition to the oxygen stoichiometry-driven structural transformation, such topotactic redox reactions accompany modifications of the oxidation state of the transition metal and, thereby, the physical properties, including magnetic and electronic ground states.[1,4,8] Therefore, another promising application of the topotactic oxides is for sensors and switches, since the electronic conductivity can be significantly modified as a consequence of the oxygen stoichiometry variation. Moreover, the material's ground state can be reset to the original state via a reversible topotactic reduction or oxidation.[1,9,10]

To utilize the oxygen sponge materials with versatile functionalities, it is highly desired to understand how the structure and physical properties evolve during the topotactic transformation. As a part of such efforts, *in-situ* diffraction and thermogravimetric studies have been conducted to reveal average structural and mass changes.[1,11,12] Such *in-situ* probes provided information on the long-range ordering and transition temperature. However, the relationship between the structure and ionic/electronic conduction has not been much understood in conjunction with the topotactic transformation. Since the topotactic transformation occurs via removal of constituent element(s), obtaining information on the changes in electronic and magnetic properties as well as crystallinity at each stage towards the transformation can provide useful information to understand the atomistic oxidation process.[13] In this regard, epitaxial oxygen sponges, $SrCoO_x$ (SCO, $2.5 \leq x \leq 3.0$), would be a promising system due to low temperature redox behavior and its coupling among structure, transport, and magnetism as were recently reported.[1,10]

In this work, we grew $SrCoO_{2.5}$ films by PLE and subsequently post-annealed in oxidizing condition without breaking vacuum to capture the each state of $SrCoO_{2.5}$ at different temperatures. X-ray diffraction, transport, and magnetization measurements of the prepared thin films show a huge structural modification and change in the electrical resistivity in close proximity to the topotactic phase transition temperature.

In order to systematically study the temperature dependent topotactic oxidation, we grew 30-nm-thick brownmillerite $SrCoO_{2.5}$ (BM-SCO) epitaxial thin films on (001) $(LaAlO_3)_{0.3}$-$(SrAl_{0.5}Ta_{0.5}O_3)_{0.7}$ (LSAT) substrates by PLE. By choosing LSAT ($a$ = 3.868 Å) as the substrate for this experiment, we could avoid being too much lattice mismatched before or after topotactic phase conversion as the substrate's lattice constant is in between the highly oxidized perovskite SCO (P-SCO, $a$ = 3.8289 Å) and BM-SCO ($a$ = 3.905 and $c$ = 3.936 Å in pseudotetragonal notation). These mismatches lead to 1% tensile strain in the fully oxidized P-SCO and -0.95% compressive strain in the BM-SCO phase. We used the same growth parameters as one can find elsewhere[1], and the film thickness was kept at 30 nm. It is also worth mentioning that with this film thickness we could maintain the full strain state on LSAT substrates unlike other substrates, including $LaAlO_3$, $SrTiO_3$, and $GdScO_3$, which revealed strain relaxation, making the full topotactic transformation difficult.

After confirming the epitaxial growth and strain state of BM-SCO thin films, we grew BM-SCO films in the identical growth condition and performed post-oxidation without breaking vacuum to avoid the surface contamination. The latter effect seems rather substantial for obtaining a low oxidation temperature. Unlike our previous approach with varying the oxygen partial pressure ($PO_2$) for oxidation,[14] we annealed as-cooled BM-SCO films at various temperatures at a fixed $P(O_2) \sim 500$ Torr. To maintain consistency and minimize other factors contributing to this annealing process, i.e. time for introducing oxygen into the chamber, we cooled as-grown BM-SCO samples below 50 °C, followed by introduction of 500 Torr of pure oxygen ($O_2$). We ramped the temperature to the annealing temperature at a rate of 10 °C/min. At each annealing temperature, we stayed for five minutes and then cooled the thin film samples in the same oxygen pressure to room temperature. In addition, we kept the annealing time (5 min.) and film thickness (30 nm) to let oxygen diffuse through entire films, which was long enough based on our optimization process. To determine the critical temperature for topotactic oxidation from the brownmillerite to the perovskite structure, we performed x-ray diffraction (XRD), temperature dependent *dc* transport measurements with a Quantum Design 14T PPMS, magnetization measurements with a Quantum Design 7 T Superconducting Quantum Interference Device (SQUID). XRD $\theta$-$2\theta$ scans were performed to check the epitaxy and phase purity as well as orientation. Also, XRD rocking curves of the 008 BM-SCO and 002 P-SCO reflections were measured to check crystallinity and structural evolution during topotactic transformation. Electronic transport measurements were done with the Van der Pauw geometry, and a proper current for the measurement has been chosen from I-V measurements at room temperature.

Figure 1 shows XRD $\theta$-$2\theta$ scans of post-annealed SCO thin films at different annealing temperatures ($T_a$). We found that the structure of samples annealed at 50 °C was essentially the same as as grown, c-axis oriented BM-SCO thin films. It is clearly seen that all the peaks are sharp. Also, no peaks other than 00*l* reflections were detected. Note that the lattice constant along the c-axis is 3.963 Å, which is expanded 0.7% from bulk BM-SCO ($c$ = 3.936

Å in pseudotetragonal notation). This expansion is caused by the substrate induced compressive strain. Note that the expected biaxial strain is 0.95% when the lattice mismatch is considered. This result implies that the film is coherently strained. The strain state was indeed clearly confirmed by XRD reciprocal space mapping (RSM, data not shown). In addition, the full width at the half maximum values ($\Delta\omega$) from rocking curve scans were about 0.05° comparable to the values of any as-grown BM-SCO films. This result confirms the annealing at 50 °C is not thermally adequate to alter the structure. After setting 50 °C as the base temperature for BM-SCO films, we performed a series of post-annealing experiments at different temperatures and checked the transition temperature, where we can find structural evolution due to the oxygen intercalation. Figure 1 shows XRD $\theta$-$2\theta$ diffraction patterns, where each sample was annealed at different temperature in between 50 and 400 °C. We observed from the samples annealed up to 175 °C that they maintained the BM-SCO phase that can be identified by the characteristic half order peaks originating from the alternating stacks of octahedral and tetrahedral layers. On the other hand, the samples annealed above 200 °C showed a clear sign of the perovskite phase, indicating the topotactic transition due to oxidation. Owing probably to the low annealing temperature, we did not observe any other impurity or intermediate phases, such as Co-O, and $Sr_6Co_5O_{15}$.[15,16] This result represents our oxidation is indeed topotactic and, thereby, bigger atoms i.e. Sr and Co are not affected but oxygen moves rather freely.[17] Note that even after this topotactic phase conversion the strain state of resulting films was retained, as was confirmed by RSM. In addition, we observed a large decrease in the *c*-axis lattice constant of P-SCO by nearly 4% as a consequence of oxidation as compared to BM-SCO, implying the role of oxygen vacancy for lattice expansion.[18]

To check how the crystallinity is affected by this oxidation process, we measured x-ray rocking curves of each annealed sample.[19] Figure 2 shows FWHM of the 008 brownmillerite peak and the 002 perovskite peak from the samples shown in Fig. 1. We only observed a negligible change in the FWHM value from brownmillerite films annealed at low temperatures ($T_a \leq 100$ °C). However, the films annealed at 150 and 175 °C clearly showed a substantial increase in the FWHM value, i.e. increased mosaicity, reflecting the fact that the microstructure of BM-SCO films was on the verge of topotactic phase change as the oxygen vacancy channels in BM-SCO were filled by oxygen. This topotactic process results in destabilization of the structure at the microscopic level modifying the bond length and angle. It is also worth pointing out that we have not seen any mixed phases from those samples.[14]

In case of the films annealed above 175 °C, however, we clearly observed not only the P-SCO phase, but greatly improved crystallinity, which was confirmed by the reduced FWHM values in rocking curve scans. This result represents that the disorder due to oxidation of BM-SCO is deeply related to a structural rearrangement in brownmillerite structure via changes in bonding angle and distance. The metastable structure on the verge of the topotactic transition becomes stabilized as the topotactic oxidation is completed as low as 200 °C. In addition, this low-temperature transformation probed by the XRD scans verifies that the energy barrier for the oxidation process is quite low.

To elucidate relation between structural transformation and electronic structure evolution, we performed *dc* transport measurements. The samples annealed above 200 °C unanimously showed a metallic behavior as shown in Fig. 3. The transport data from films annealed above 200 °C are clear evidence for stabilization of the perovskite phase, consistent with the XRD results. While the films annealed at 150 and 175 °C showed still an insulating behavior, the resistivity values are quite reduced as compared to the as-grown brownmillerite film. This result indicates a possible change of the oxygen stoichiometry as the increased oxygen content in BM-SCO yields a valence change of Co ions from $Co^{3+}$ to $Co^{4+}$. As SCO is known to be a *p*-type conductor based on thermoelectric measurements[1], oxygen intercalation can result in an increase of conducting charges. Furthermore, in order to check how the thermally induced activation energy changes in this brownmillerite structure, we calculated the thermal activation energy ($E_a$) of non-metallic samples, as summarized in the inset of Fig 3. The as-grown BM-SCO film on LSAT revealed $E_a$ = 140 meV. This value is lower than that of BM-SCO on STO ($E_a$ = 190 meV) and bulk BM-SCO (240 meV).[14,20] We attribute the lower activation energy to the reduced in-plane hopping distance by biaxial compressive strain in particular for BM-SCO/LSAT, since the crystal structure is closely coupled with electronic properties in transition metal oxides.[10,21,22], In addition, the inset of Fig. 3 clearly shows a drastic decrease in $E_a$ as it reaches to the insulator-to-metal transition via local oxidation. The sharp reduction in $E_a$ also indicates that the sample annealed at 175 °C might be useful for technical applications, such as solid oxide fuel cells and catalysts, as the metastable oxygen stoichiometry plays a critical role for their operations. Based on XRD and transport data, we estimate the oxygen concentration ($SrCoO_x$) of this sample to be *x* ~ 2.75 or slightly lower, in which the magnetic and electronic ground states are known to be significantly modified.[14,23]

We further checked the magnetic properties of the films to observe the magnetic transition upon the topotactic oxidation. Similar to the XRD results, we observed an obvious change of the magnetic ground state, when the macroscopic crystal structure changed from brownmillerite to perovskite. Note that any evidence of low temperature ferromagnetism due to local oxidation from the film annealed at 150 and 175 °C was not found. As shown in Fig.

3 and Fig. 4, the electronic transport measurements with P-SCO films annealed above 200 $^{o}$C revealed clear metallicity and ferromagnetism. Despite the decreased magnetic critical temperature in the oxidized thin films compared to that of ozone-grown P-SCO[1], topotactically oxidized P-SCO is a clear ferromagnetic metal below ~ 220 $^{o}$C. Lastly, we observed a slight upturn of the resistivity from P-SCO films oxidized at 400 $^{o}$C (see Fig. 3). This could be attributed to the fact that the increased thermal hopping shed lattice oxygen by overcoming the oxygen background pressure, yielding a less oxidized state as one can find also reduced magnetization from the sample annealed at 400 $^{o}$C as shown in Fig. 4. This result indicates the thin films annealed at 400 $^{o}$C may have an oxygen concentration lower than $x = 2.90$ due to local oxygen vacancy formation.

In conclusion, by investigating the structure and associated transport properties in epitaxial SrCoO$_x$ oxygen sponges, we clearly observed enhanced structural disorder during the topotactic phase transition (~175 $^{o}$C). While still insulating and non-ferromagnetic, the oxygen sponge near the topotatic transition revealed a greatly reduced thermal activation energy due to oxygen intercalation. Above the transition temperature, the films revealed a clear transition into the perovskite phase, recovering the high crystallinity as good as what was found for as grown brownmillerite epitaxial thin films. This result indicates that the topotatic phase control via oxygen stoichiometry is a fascinating route to not only obtaining high quality crystals, but tuning the physical properties without chemical doping of heavy elements. Moreover, due to the conserved crystallinity even after structural phase transition, the phase of oxygen sponges can be repeatedly reversed, which is useful for the development of novel electrochemical devices, i.e. solid oxide fuel cell, rechargeable batteries.

**Acknowledgments:** We greatly appreciate J. Petrie for technical assistance. This work was supported by the U.S. Department of Energy, Office of Science, Basic Energy Sciences, Materials Sciences and Engineering Division.

**Fig. 1**. XRD $\theta$-$2\theta$ scans of SrCoO$_x$ films on (001) LSAT substrates. The films were postannealed at temperatures up to 400 °C to observe the topotatic transition from the brownmillerite SrCoO$_{2.5}$ to the perovskite SrCoO$_{3-\delta}$. The perovskite phase was formed by annealing at 200 °C or higher.

**Fig. 2.** FWHM values from XRD rocking curve scans using the 008 reflection for brownmillerite films and the 002 reflection for perovskite films. A clear increase in FWHM is observed on the verge of topotatic conversion (~175 °C) from BM-SCO to P-SCO . The inset shows a rocking curve for the P-SCO film annealed at 200 °C.

**Fig. 3.** Temperature-dependent resistivity data for SCO thin films annealed at different temperatures. Samples annealed at 150 °C or lower revealed a clear insulating behavior indicating that oxygen intercalation is limited. Whereas, a clear metallic ground state was observed from samples annealed at 200 °C or higher. Interestingly, even though the crystallinity is the worst among the samples we investigated, more or less semiconducting behavior was observed from the sample annealed at 175 °C due to a partial oxygen intercalation. The inset shows the thermal activation energy ($E_a$) of insulating films. A clear decrease in $E_a$ was observed as a consequence of topotactic oxygen intercalation. Note that the resistivity data of the SCO thin film annealed at 300 °C is adopted from Ref. 1.

**Fig. 4.** Magnetic characterization of the SCO thin films annealed at 50, 175, 200, 300, and 400 °C: (a) Magnetization as a function of temperature with 0.1 T, and (b) magnetization hysteresis at 10 K. The film annealed at 400 °C showed a lower

magnetization and a higher resistivity than those from 200 and 300 $^{o}$C annealed samples, reflecting the reduced oxygen concentration due to the loss of oxygen by increased thermal hopping of oxygen.

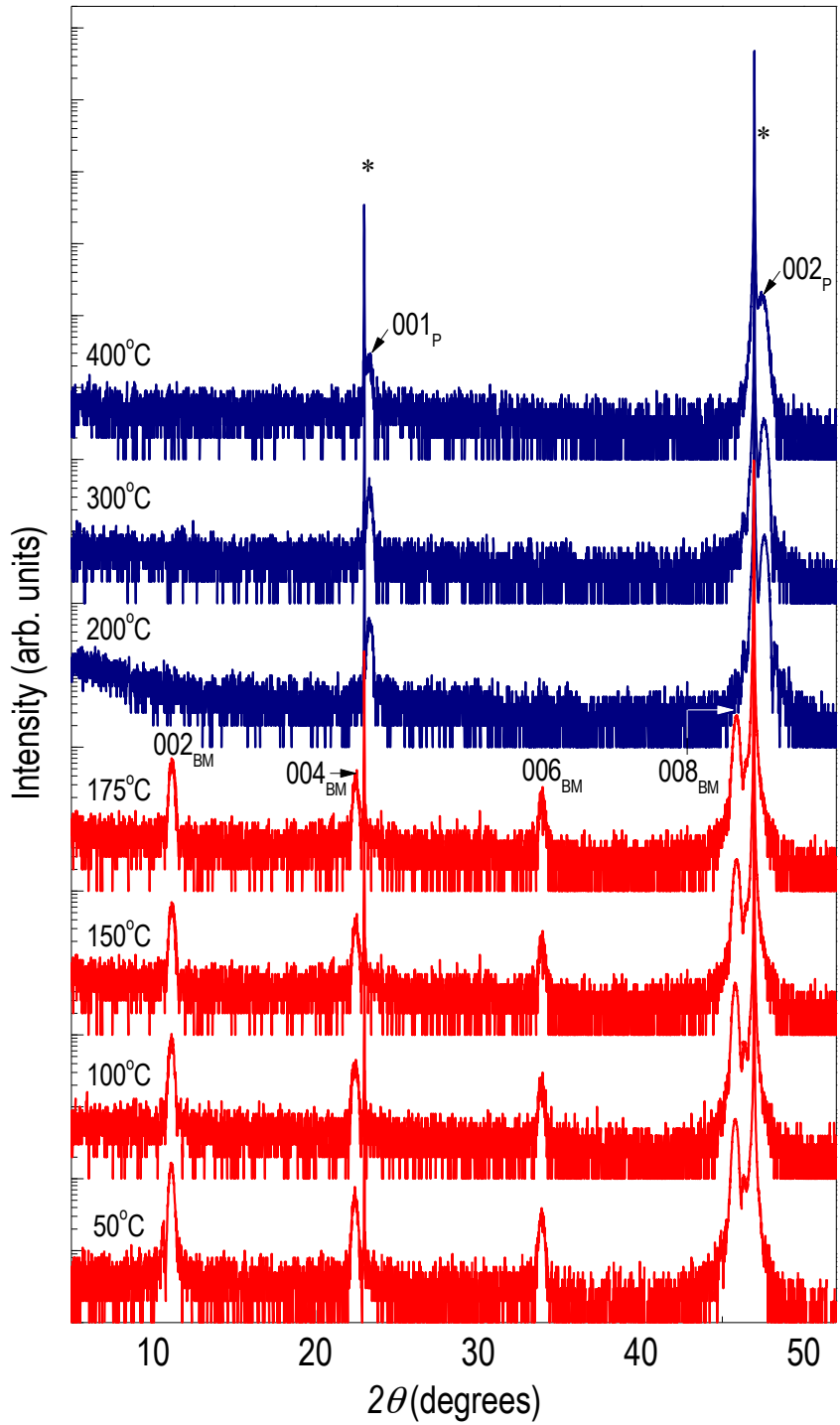

Fig. 1

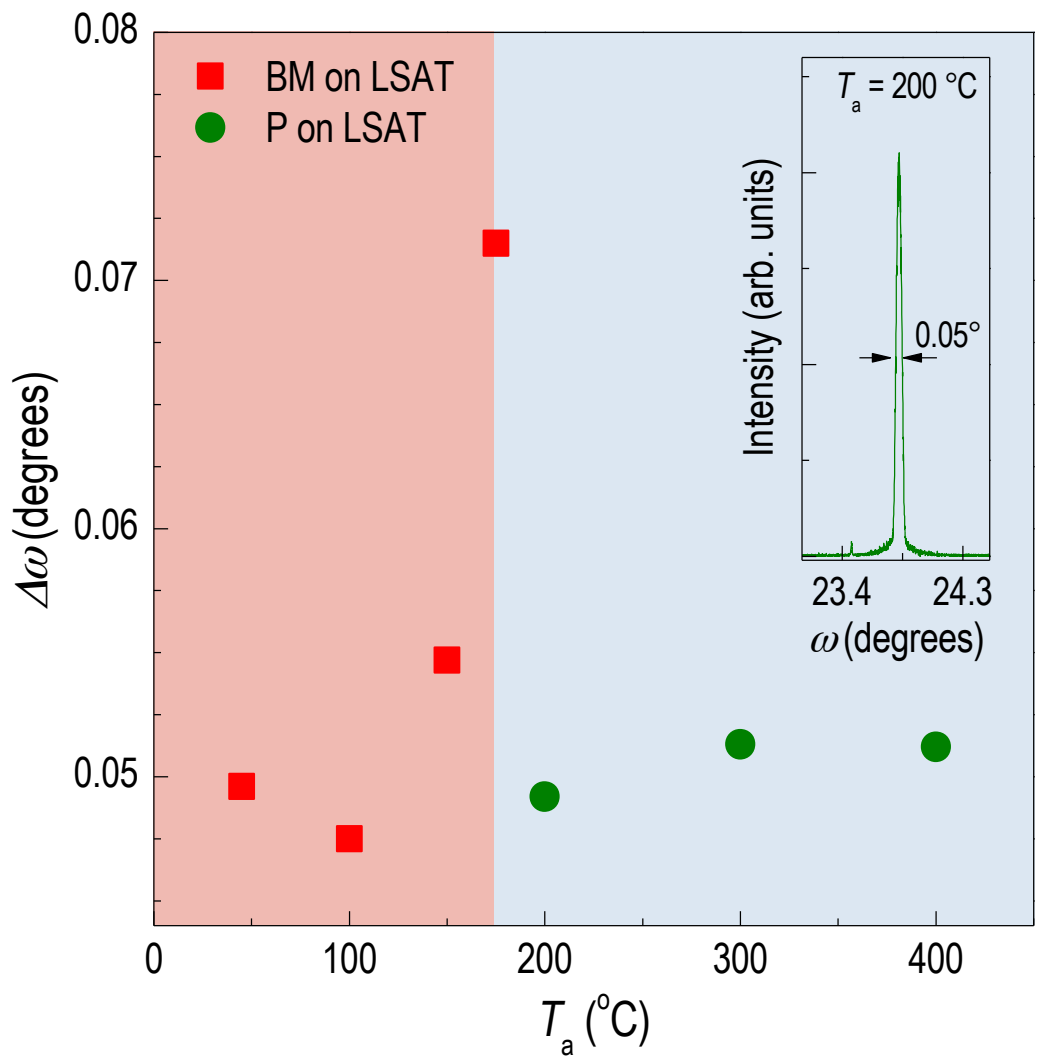

Fig. 2

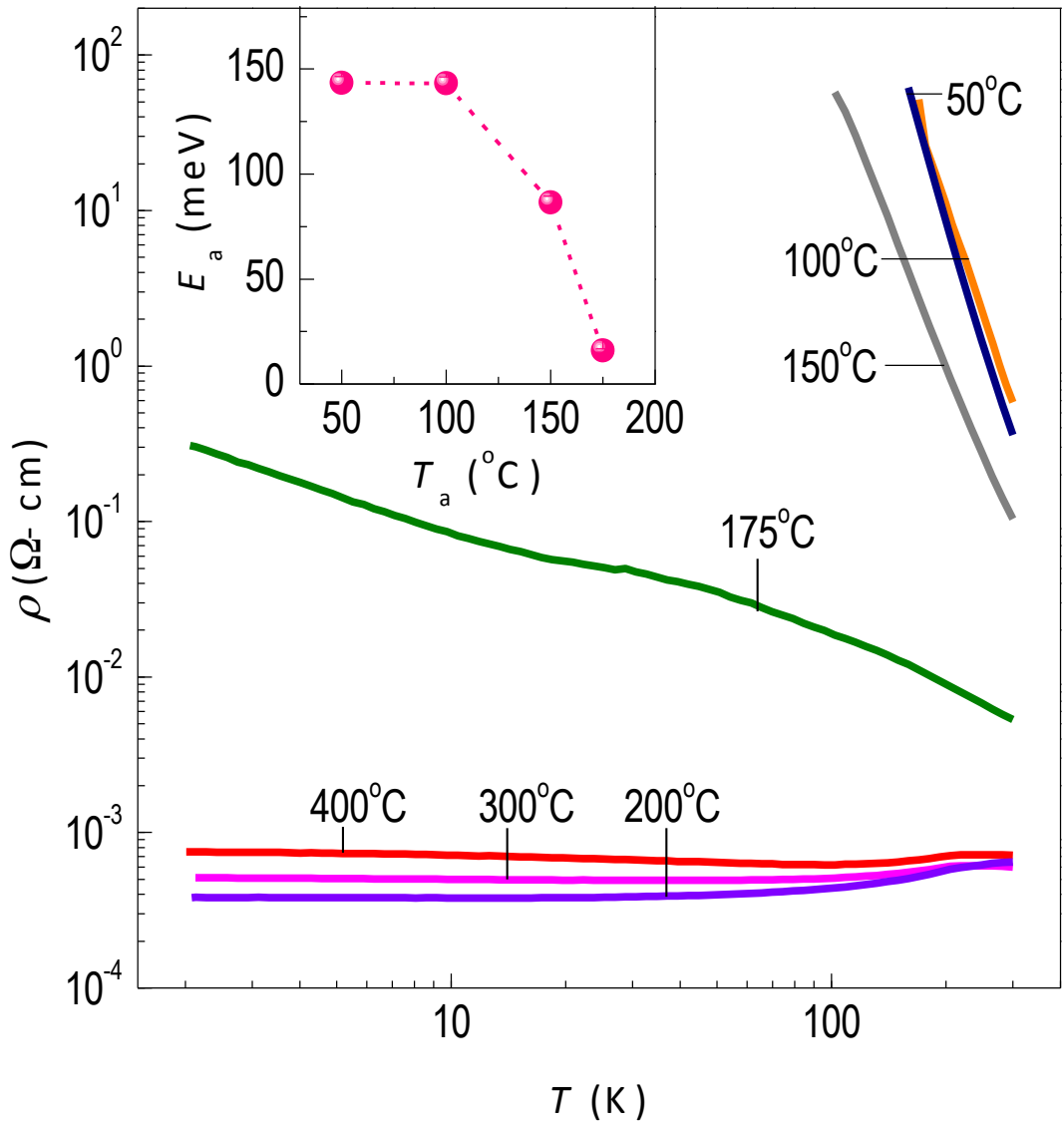

Fig. 3

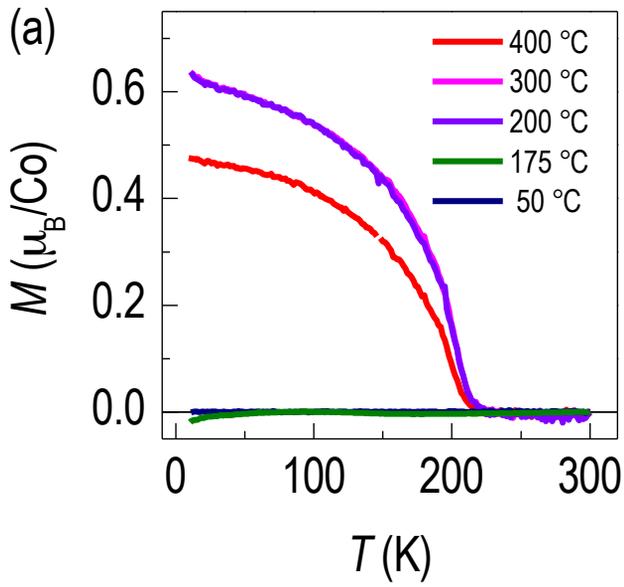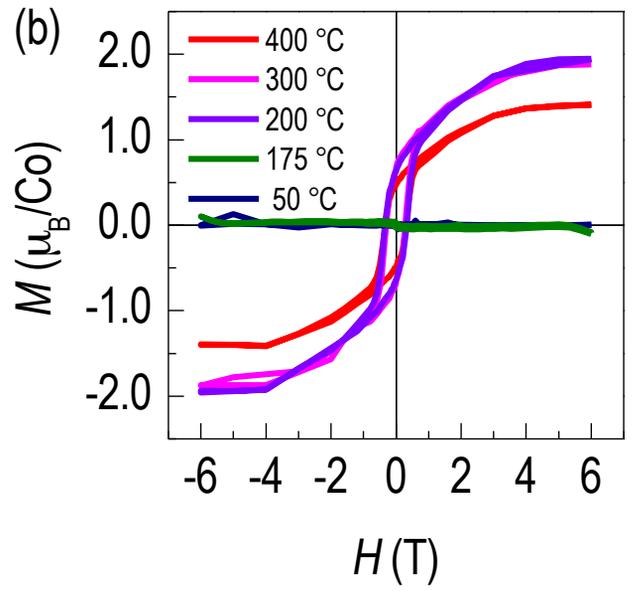

Fig. 4